\documentclass[journal]{IEEEtran}
\usepackage{cite}
\usepackage{amsmath,phonetic,graphicx}
\usepackage{textcomp,algorithmic,multirow,hhline}
\usepackage{amsfonts}
\usepackage[table]{xcolor}

\begin{document}
\title{Deep Speech Enhancement for Reverberated and Noisy Signals using Wide Residual Networks}
\author{\IEEEauthorblockN{Dayana Ribas, Jorge Llombart, Antonio Miguel, Luis Vicente}\\
\IEEEauthorblockA{\textit{ViVoLab, Aragon Institute for Engineering Research (I3A)} \\
\textit{University of Zaragoza, Spain}\\
}
}

\maketitle

\begin{abstract}
This paper proposes a deep speech enhancement method which exploits the high potential of residual connections in a wide neural network architecture, a topology known as Wide Residual Network. This is supported on single dimensional convolutions computed alongside the time domain, which is a powerful approach to process contextually correlated representations through the temporal domain, such as speech feature sequences. We find the residual mechanism extremely useful for the enhancement task since the signal always has a linear shortcut and the non-linear path enhances it in several steps by adding or subtracting corrections. The enhancement capacity of the proposal is assessed by objective quality metrics and the performance of a speech recognition system. This was evaluated in the framework of the REVERB Challenge dataset, including simulated and real samples of reverberated and noisy speech signals. Results showed that enhanced speech from the proposed method succeeded for both, the enhancement task with intelligibility purposes and the speech recognition system. The DNN model, trained with artificial synthesized reverberation data, was able to deal with far-field reverberated speech from real scenarios. Furthermore, the method was able to take advantage of the residual connection achieving to enhance signals with low noise level, which is usually a strong handicap of traditional enhancement methods.    
\end{abstract}


\section{Introduction}
\IEEEPARstart{T}{he} development of robust methods in speech processing has attracted a lot of attention. Robust processing addresses issues raised by actual usage scenarios to maintain, or even improve, the system performance. There can be methods for robust processing in any area of the system pipeline, either replacing some stage in the system, e.g. robust feature extraction, or as an additional method in some critical step, e.g. feature normalization for channel effects compensation. Speech enhancement methods appear in the pre-processing stage, acting directly on the speech signal. This could target two different purposes, either improve speech intelligibility or feeding a robust speech processing system, such as speech, speaker or language recognition systems. The best scenario would be to design a speech enhancement method that achieves improvement in both goals at the same time, even though it can be very challenging \cite{Loizou2013}.

Speech enhancement has been the target of much research efforts for several decades. Indeed, a lot of advances in the understanding of environmental acoustic distortion of speech signals have been seen during this period, and numberless interesting proposals have been presented attempting more accurate enhancement methods. The problem and related solutions differ whether the signals are single or multiple channels. Speech enhancement has been traditionally performed using statistical methods, giving place to established approaches such as spectral filtering techniques, e.g. Spectral Subtraction \cite{Boll1979} and Wiener filtering \cite{Lim1979}. Also clean speech estimators, such as the Minimum-Mean Square Error (MMSE) estimator \cite{Ephraim1984} and its log-spectral amplitude estimator \cite{Ephraim1985}, which has been the inspiration for many speech estimation methods, e.g. Multiplicatively-Modified Log-Spectral Amplitude (MM-LSA)\cite{Malah1999}, Optimally-Modified Log-Spectral Amplitude (OM-LSA) \cite{Cohen2001}, Minima Controlled Recursive Averaging (MCRA) \cite{Cohen2003}, among others. In addition to the spectral magnitude, a research line using phase-aware methods has developed very interesting alternatives \cite{Mowlaee2015,Mowlaee2016}. However, despite statistical algorithms have settled as state of the art for many decades, they have some important drawbacks that limit their performance, especially for non-stationary noises. For example, they make unrealistic assumptions, e.g. the uncorrelated nature of spectral coefficients in a speech frame, when spectral coefficients are actually correlated in different frequencies as well as at different time instants \cite{Cohen2008}. They also require a running estimate of noise and clean speech variances, however, such estimates are typically poor for highly non-stationary noisy samples. 

On the other hand, learning based methods, as Non-negative Matrix Factorization (NMF) \cite{Mohammadiha2013,Fan2014} and Neural Networks \cite{Wan1999}, have also been explored for speech enhancement applications. The paradigm of data-driven methods might be a suitable solution to handle the complex process of acoustic speech distortion. Lately, many Deep Neural Network (DNN) architectures have been developed with promising results. From autoencoders to feed-forward deep networks, several DNN architectures have been tested on the task of speech enhancement. 

This paper presents an overview of the DNN state of the art in speech enhancement and proposes a method based on one of the most recent and promising deep learning architectures in speech processing: the Wide Residual Neural Networks (WRN). The manuscript delineates the development of deep speech enhancement, passing through many established DNN architectures and discussing their application to the speech enhancement task. Next, a novel WRN-based speech enhancement method is introduced. In this framework, we discuss that DNN paradigm outperforms the limitations of the traditional statistical-based approaches by building estimators that well approximate the complex scenarios caused by the ubiquitous acoustic distortion in the real world.         

In Section \ref{sec:dnn} we outline several DNN-based solutions applied to the problem of speech enhancement. Section \ref{sec:speechenhancement} presents the WRN architecture and the characteristics that make it interesting for the task of denoising speech signals. In the rest of the paper a proposal based on WRN is introduced. Section \ref{sec:exp} describes the experimental setup designed for testing the proposal performance through speech quality measures and the result of an Automatic Speech Recognition (ASR) system. Section \ref{sec:res} discusses the obtained results and finally section \ref{sec:conc} concludes the paper.

\begin{figure*}[h!]
\centerline{\includegraphics[width=180mm]{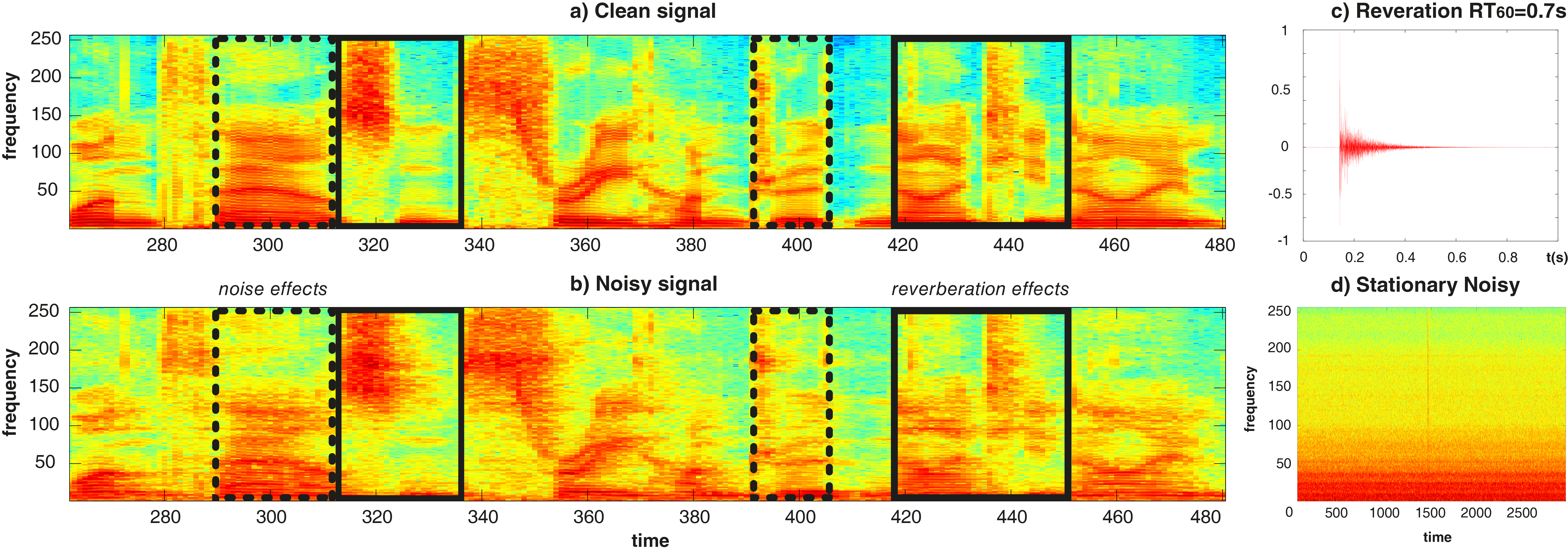}}
\caption{{\it a) Clean speech signal, b) Signal a) affected by noise in d) and reverberation in c).}}
\label{fig:noisereverb}
\end{figure*}

\section{Deep speech enhancement}
\label{sec:dnn}
In the last few years, a flood of DNN solutions has overspread many areas of research development, including speech processing. The high capacity of the deep learning approaches for finding underlying relations on the data, and providing substantial representations from them, has attracted a lot of attention of the community. The data-driven paradigm behind deep learning is suitable for modeling the relationship between noisy and clean data. This is how the denoising autoencoder (DAE) emerged, which is a combination of an encoder function that converts the input data (corrupted data) into a different representation, and a decoder function that converts the new representation back into the original format (the clean ``repaired'' data) \cite{Vincent2008}. Although this concept was first applied in the field of computer vision for handling noisy images\footnote{The proof of concept was initially performed on the MNIST digit classification problem}, it is also suitable for the problem of speech enhancement. The first approaches using DAE for speech enhancement appeared from 2012 on \cite{Lu2012,Xia2013,Lu2013,Araki2015}, and recently it has been presented a variational approach \cite{Bando2018}. The intermediate representation obtained from the autoencoder can also be used as a robust parametrization for the noisy/reverberated speech in order to feed a DNN for speech or speaker recognition \cite{Feng2014, Mimura2015, Heymann2015, Plchot2016}. This is also an alternative way to enhance speech, which has been similarly applied using Restricted Boltzmann Machines (RBM) \cite{Xu2014}.

Beyond DAE, the problem of speech enhancement has been addressed with several deep learning model architectures, often imported from previously successful experiences in computer vision. The availability of deep speech enhancement solutions has grown fast during the last few years. Many interesting solutions based on DNN -- i.e. general feed-forward deep networks -- have been proposed \cite{YongXu2015,Tu2017,Yang2018,Morten2018,Zhao2018,Karjol2018}, considering either the feature-mapping strategy, which consists on learning a non-linear mapping function between corrupted and clean speech signals, or the mask-learning strategy, which computes a ratio mask and applies it to the corrupted speech to mask the distortion and recover the clean speech. 

\subsection{Recurrent connections}
More complex architectures, such as Recurrent Neural Networks (RNN) and the associated Long Short-Term Memory (LSTM) alternative, were also adopted for feature enhancement in robust speech recognition \cite{Maas2012,Wollmer2013,Weninger2015,Chen2017,Gao2018}. They capture the temporal nature of speech using recurrent connections, achieving robust representations despite speech corruption. Then, resultant representations can be used in robust speech recognition with improved system performance. However, results have indicated the generalization capacity of RNN would weaken if it is trained on limited noise types. 

\subsection{Convolutional networks and residual connections}
Architectures based on Convolutional Neural Networks (CNN) are capable of exploiting local patterns in the spectrogram from both, frequency and temporal domains. The effect of noise and reverberation appears as a perturbation of the signal spectral shape extended through a certain time-frequency area. Figure \ref{fig:noisereverb} shows one example of these acoustic distortions. Note how consecutive time-frequency bins show the correlation in a context, due either to the natural structure of the speech signal or to the distortion patterns. CNN-based architectures effectively deal with this characteristic of the speech signal structure, therefore they are very powerful for speech enhancement purposes. Previous works have explored CNN architectures for speech enhancement tasks \cite{Fu2016, Park2017}. 

CNN has also appeared combined with recurrent blocks to further model the dynamic correlations among consecutive frames \cite{HanZhao2018}. The incorporation of residual connections brought more potential to the CNN approach. They make use of shortcut connections between neural network layers, allowing to handle deeper and more complicated neural network architectures, with fast convergence and a small gradient vanishing effect. Thus, they are be more expressive and provide more detailed representations of the underlying structure of the corrupted signal, resulting in more accurate enhanced speech. For the time being, residual connections have been barely used in speech enhancement \cite{Chen2017}. 

\subsection{Adversarial networks, Wavenet and more}
Moreover, the combination of generative and discriminative modeling has been used for speech enhancement through the Generative Adversarial Network (GAN) \cite{Pascual2017,Michelsanti2017,Soni2018,Donahue2018}. GAN proposes an adversarial scheme where one-sided network model generates enhanced speech proposals, while the other-side discriminative model is trained to distinguish these from the clean reference speech. This way, the full network architecture finely adjusts the parameters in order to obtain a final clean "repaired" speech signal. 

Other interesting DNN architectures, such as Wavenet \cite{Qian2017}, have also addressed the problem of speech enhancement from a generative point of view. This way they are able to predict the output signal sample by sample, such that from a corrupted signal, the clean speech can directly be recovered. In addition, recent advances in probability density function definition using neural networks have made possible to manage the uncertainty of the prediction and, more importantly, they have accelerated its computational performance which makes it a solid alternative to spectrum based systems in the future.

Additionally to the network architecture designs, several techniques have been proposed to improving the DNN-based speech enhancement system. The global variance equalization alleviates the over-smoothing problem of the regression model. The dropout and noise-aware training strategies contribute to improve the generalization capability of DNN to unseen noise conditions \cite{YongXu2015}.

\section{Wide Residual Networks for speech enhancement: Proposal}
\label{sec:speechenhancement}
As mentioned in the previous section, residual connections have shown to be able to scale up the DNN to thousands of layers with improving performance. An alternative to this topology is decreasing the depth and widening the layers in order to gain more convolution channels. This trade-off allows keeping the advantages of residual connections and accelerates the training process. This architecture is called Wide Residual Network (WRN) \cite{Zagoruyko2017}. 

The speech enhancement proposal of this paper consists of a denoising DNN with multiple input sources based on a WRN architecture. The proposed network conveys several Wide Residual Blocks (WRB) that follow the main path -- two single dimensional convolutional layers, Batch Normalization (BN) and a Rectified Linear Unit (ReLU) non-linearity -- in parallel to a residual path connection between input and output. The residual connection provides a very useful mechanism for the enhancement purpose since the signal has available a linear shortcut for adding or subtracting corrections. The non-linear main path has the potential to improve the signal or leave it unchanged as needed.

\begin{figure}[h!]
\centerline{\includegraphics[width=80mm]{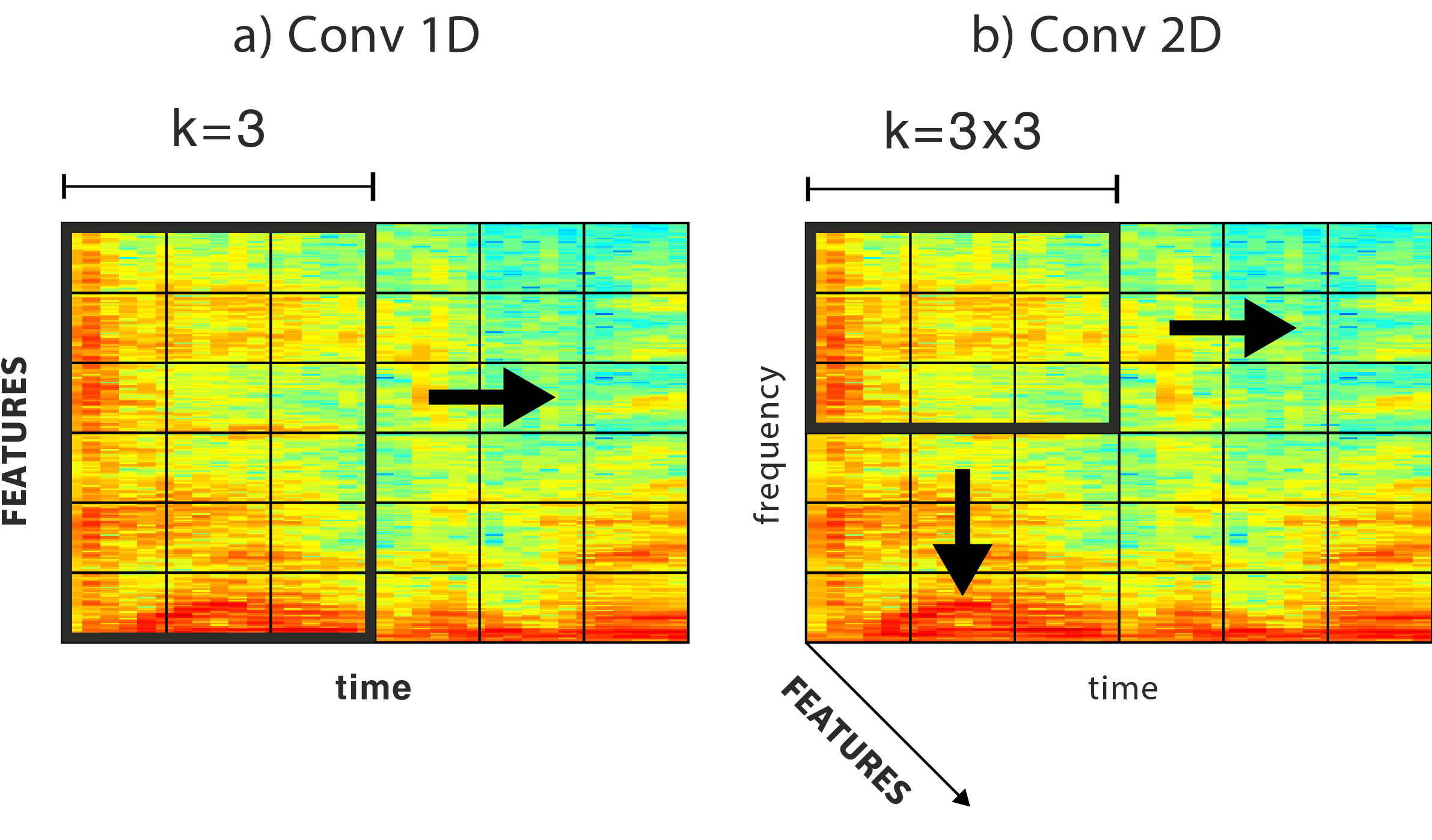}}
\caption{{\it Scheme of convolutions in a CNN.}}
\label{fig:conv}
\end{figure}

Convolutional layers in this network are computed through the single temporal dimension, so we will call them Conv1D. This approach comes from image processing, hence, in speech processing some works have processed the spectrum as an image. In this case, convolutions are computed through both, the temporal and frequency dimensions, i.e., two-dimensional convolutions (Conv2D). The main difference between Conv1D and Conv2D is the information considered by the convolutional filter, i.e. the Conv2D filter moves both, through temporal and frequency domains, while the Conv1D filter moves only along the temporal dimension. This way, the Conv2D filter considers the value of the spectrum at any time-frequency location $x_{t,d}$ as features, where $t \in T$ are time frames and $d \in D$ are frequency bands. On the other hand, the feature vector for the Conv1D filter contains the full frequency band values of the involved time frames $x_t$, so the input is $X=[\vec{x}_1, ..., \vec{x}_t, ..., \vec{x}_T]$ where $\vec{x}_t \in R^D$ is the feature vector. Figure \ref{fig:conv} depicts this process. 

Speech is a single dimensional signal that is usually processed using sliding windows. Therefore, we consider Conv1D is better suited for handling speech signals, because at every time instant it provides a full-frequency feature vector that completely describes the sample. 

Convolutional layers are usually described by the kernel dimension $k$ which is the number of samples used in the filter. This case we use $k=3$. The output of a convolutional layer $(L)$ with $k=3$ can be described as $\vec{o}_t^{L+1}=f_{L}(\vec{x}_{t-1}^{L}, \vec{x}_t^{L}, \vec{x}_{t+1}^{L})$ with $\vec{o}_t^{L+1} \in R^{D^{L+1}}$. $D^{L+1}$ usually is known as the number of neurons in the convolutional layer, but also as the number of output channels regarding the image context. When $D^{L+1} > D^{L}$ the number of channels increases, i.e. the network structure widens. At this point, convolutional layers can be viewed as a special case of Time Delay Neural Networks (TDNN) \cite{Povey2015} where some future and past samples are concatenated to each sample as the input of a neuronal layer. 

\begin{figure*}[h!]
\centerline{\includegraphics[width=160mm]{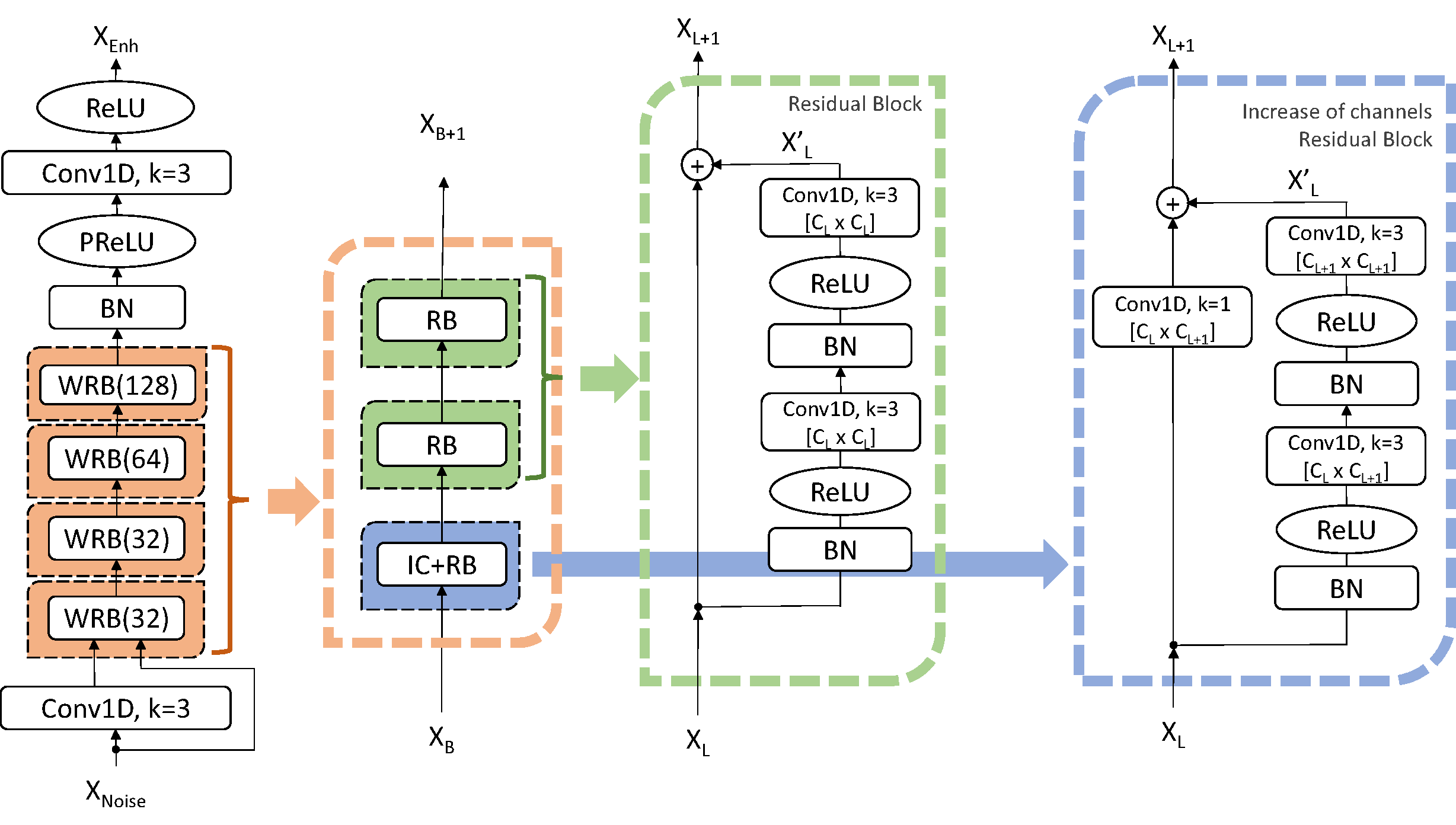}} 
\caption{{\it WRN architecture proposed. From left to right there is the composition of the network blocks, with $C_{L}$ the number of channels in the layer $L$.}}
\label{fig:WRNDAE4}
\end{figure*}

Conv1D can easily incorporate different sources of information by concatenating their channels. This way, the network is fed with feature vectors that are obtained from different speech representations in the time-frequency domain: the magnitude of the Fast Fourier Transform (FFT), Mel filterbank, and cepstrum. Each of them provides an alternative view of the speech signal that enriches the characterization of the corrupted signal structure. To address the mismatch between the analysis window size and the channel of the filterbank, feature vectors are computed in three different windows sizes. 

The network architecture proposed processes input features with a first convolutional layer followed by four WRB. The first WRB processes the output of the first convolutional layer and also the input of it. Following the WRBs there is a BN stage, a non-linearity (PReLU: Parametric ReLU), and finally, another convolutional layer with a ReLU, to reduce the number of channels to 1 and obtain the enhanced FFT for each signal. In each of the WRBs we increase the number of channels used to get the outputs. The widen operation is done in the first convolution of the first residual block of each WRB. In order to compute the residual connection as a sum operation, the number of channels in the straight path and in the convolution path has to be the same. Therefore, when the number of channels is increased, a Conv1D $k=1$ is added. This can be interpreted as a position wise fully connected layer to adjust the number of channels with the channels in the convolutional path. Figure \ref{fig:WRNDAE4} illustrates the proposed architecture.

\section{Experimental setup}
\label{sec:exp}
The experimental framework developed in this work is based on the REVERB Challenge task\footnote{http://reverb2014.dereverberation.com}. To test the performance of the speech enhancement method, two experimental lines are conducted: 1) The computation of speech quality measures and 2) the assessment of a speech recognition system performance. The dataset specifications are described in the next section, followed by the quality measures and the ASR system used in the experiments.

\subsection{Datasets}
The approach was tested on the official development and evaluation sets of the REVERB Challenge \cite{Challenge2013}. The dataset has simulated speech from the convolution of WSJCAM0 Corpus \cite{WSJCAMO} with three measured Room Impulse Responses (RIR) and added stationary noise recordings from the same rooms. Besides, it has real recordings, acquired in a reverberant meeting room from the MC-WSJ-AV corpus \cite{MC-WSJ-AV}. It also has a multicondition dataset with 7861 utterances from 92 speakers, which is artificially distorted speech data similar to the simulated dataset. The full REVERB is sampled at 16 kHz. Impulse responses and noise types are chosen randomly, with equal probability. We also tested on the freely available version of DIRHA dataset\footnote{http://shine.fbk.eu/resources/dirha-ii-simulated-corpus}. It contains simulated acoustic sequences of real background noise recorded in five rooms of an apartment -- living room, bathroom, bedroom, kitchen, and corridor -- with various localized acoustic events superimposed, i.e. speech and noises. Original data at 48 kHz were re-sampled at 16 kHz for processing. See table \ref{tablereverb} for details.

\begin{table} [h!]
\caption{\label{tablereverb} {\it Description of REVERB dataset used for computing speech quality measures and ASR system performance. $Distance$: Speaker-microphone distance, $RT_{60}$: Reverberation Time}}
\vspace{1mm}
\centerline{
\begin{tabular}{|c|c|c|} 
\hline
\cellcolor{gray!35}\textbf{Dataset} & \cellcolor{gray!35}\textbf{REVERB SimData} & \cellcolor{gray!35}\textbf{REVERB RealData} \\ \hline 
Signals & dev: 1484 / eval: 2176 & dev: 179 / eval: 372 \\ \hline 
Speakers & dev: 20 / eval: 20 & dev: 5 / eval: 10 \\ \hline 
Speech type & \multicolumn{2}{c|}{Read} \\ \hline 
Interface & \multicolumn{2}{c|}{Microphone} \\ \hline 
Distance (cm) & near: 50, far: 200 & near: 100, far: 250  \\ \hline
Room size & 1) small, 2) medium, 3) large & - \\ \hline
$RT_{60}$ (s) & 0.25, 0.5, 0.7 & 0.7 \\ \hline 
Noise type & \multicolumn{2}{c|}{Stationary (air conditioner)} \\ \hline
SNR (dB) & \multicolumn{2}{c|}{20} \\ \hline
\cellcolor{gray!35}\textbf{Dataset} & \multicolumn{2}{c|}{\cellcolor{gray!35}\textbf{DIRHA}} \\ \hline 
Signals & \multicolumn{2}{c|}{263} \\ \hline 
Speech type & \multicolumn{2}{c|}{Spontaneous} \\ \hline 
Interface & \multicolumn{2}{c|}{Microphone} \\ \hline 
Distance (cm) & \multicolumn{2}{c|}{Random inside the rooms} \\ \hline
Rooms & \multicolumn{2}{c|}{livingroom, bathroom, bedroom, kitchen, corridor} \\ \hline
$RT_{60}$ (s) & \multicolumn{2}{c|}{0.74, 0.75, 0.68, 0.83, 0.60} \\ \hline 
Noise type & \multicolumn{2}{c|}{Home background noises} \\ \hline
\end{tabular}}
\end{table}

\begin{table*} [h!]
\caption{\label{tablednntraining} {\it Description of dataset used for DNN training.}}
\vspace{1mm}
\centerline{
\begin{tabular}{|c|c|c|c|c|c|c|} 
\hline
\cellcolor{gray!35}\textbf{Dataset} & \cellcolor{gray!35}\textbf{Timit} & \cellcolor{gray!35}\textbf{Librispeech} & \cellcolor{gray!35}\textbf{RSR2015} & \cellcolor{gray!35}\textbf{TedLium} & \cellcolor{gray!35}\textbf{Voxforge} & \cellcolor{gray!35}\textbf{Commonvoice} \\ \hline 
Signals & 6299 & 292329 & 196688 & 56704 & 58961 & 195607 \\ \hline 
Speakers & 630 & 2484 & 300 & 698 & 2521 & 20000 \\ \hline 
Speech type & \multicolumn{2}{c|}{Read} & Prompted speech & Talk & \multicolumn{2}{c|}{Prompted speech} \\ \hline 
Interface & \multicolumn{2}{c|}{Microphone} & Cell, Tablet & Audio from Video & \multicolumn{2}{c|}{Computer, Tablet, Cell} \\ \hhline{=======} 
\multicolumn{7}{|c|}{\cellcolor{gray!35}\textbf{Data augmentation}} \\ \hline
Room size \textit{[x,y,z]} (m) & \multicolumn{2}{c|}{small: [2-6, 2-6, 2.5-3.5]} & \multicolumn{2}{|c|}{medium: [6-15, 6-15, 3-5]} & \multicolumn{2}{|c|}{large: [10-20, 10-20, 4-6]} \\ \hline
$RT_{60}$ (s) & \multicolumn{2}{c|}{0.05 - 0.3} & \multicolumn{2}{|c|}{0.1 - 0.5} & \multicolumn{2}{|c|}{0.6 - 0.8} \\ \hline 
Distance (cm) & \multicolumn{2}{c|}{0 - 400} & \multicolumn{2}{|c|}{0 - 900} & \multicolumn{2}{|c|}{0 - 1000}  \\ \hline
Noise type & \multicolumn{6}{c|}{Stationary and non-stationary noise recordings, including music and speech} \\ \hline
SNR (dB) & \multicolumn{6}{c|}{10 - 20} \\ \hline
\end{tabular}}
\end{table*}

For training the DNN we used 16 kHz sampled data from the following datasets: Timit\footnote{https://catalog.ldc.upenn.edu/LDC93S1}, Librispeech\footnote{http://www.openslr.org/12/}, RSR2015\footnote{https://www.accelerate.tech/innovation-offerings/ready-to-sign-licenses/rsr2015-overview-n-specifications}, Tedlium\footnote{http://www.openslr.org/7/}, Commonvoice\footnote{https://voice.mozilla.org/en/data}, and Voxforge\footnote{http://www.voxforge.org/}. This data was augmented by adding randomly selected RIR \cite{allen1979}, noises from Musan dataset \cite{musan2015}, and scaling the time axis at the feature level. 
Table \ref{tablednntraining} shows further details on the DNN training dataset and data augmentation resources. 

\subsection{DNN configuration}
The front-end starts segmenting speech signals in 25, 50, and 75 ms Hamming window frames, every 10 ms (15, 40, 65 ms length). For each frame segment, three types of acoustic feature vectors are computed and stacked, to create a single input feature vector for the network: 512-dimensional FFT, 32, 50, 100-dimensional Mel filterbank, and cepstral features (same dimension of the corresponding filterbank). Finally, each feature vector is normalized by mean and variance. 

Speech enhancement was performed through a WRN architecture \cite{He2016,Zagoruyko16} with Conv1D of kernel size 3. All these was built on the pytorch toolkit. Input features for training the network, described in the previous paragraph, were generated on-the-fly, operating in contiguous vector blocks of 200 samples, so that convolutions in the time axis can be performed. The network uses four WRN blocks with a widen factor of 8. AdamW algorithm was used to train the network \cite{Kingma2015,Loshchilov2017} and PReLUs \cite{He2015} as parametric non-linearity. The cost function for a segment of contiguous frames is based on the Mean Square Error (MSE) of the logarithmic FFT of the input and the clean (training mode) / enhanced (evaluation mode) output (eq. \ref{eq:cost}):

\begin{equation}
    \label{eq:cost}
    cost = \frac{1}{T} \sum_{t=1}^{T} \sum_{f=1}^{D} (\log \vec{\hat{x}}_{t,f} - \log \vec{x}_{t,f})^2 
\end{equation}

Finally, to produce the enhanced sample the fourth WRN blocks were connected to a batch normalization layer \cite{Ioffe15}, a PReLU, a position wise fully connected, and a final Conv1D with ReLU to reduce the number of channels to 1 and finally obtaining the enhanced FFT for each signal.

To get enhanced waveforms from the FFT produced by the DNN, we coupled the spectral phase of the corresponding noisy signal to the enhanced magnitude spectrum. The full spectrum was inverted through an IFFT to obtain the time domain representation. This was windowed compensated and finally overlap-added for converting the separated frames to a temporal sequence. 

\subsection{Speech quality measures}
In order to measure the speech enhancement performance the enhanced signals were compared to the original clean dataset using the following objective speech quality measures:

\begin{itemize}
\item Cepstrum distance (CD) \cite{CDmeasure}: Distance among features in the cepstral domain. It is computed using the observed/enhanced signal and the clean reference. The closer the target feature to the reference, the better the quality. Therefore smaller CD values indicate better speech quality.
\item Log-likelihood ratio (LLR) \cite{Loizou2011}: Represents the degree of discrepancy between smoothed spectra of the target and reference signals, computed over the Linear Prediction Coefficients. Smaller values indicate better speech quality. 
\item Speech-to-reverberation modulation energy ratio (SRMR) \cite{SRMRmeasure}: Measures the energetic relation between speech and reverberation in the modulation spectral domain. In this case, larger values indicate better speech quality.
\item Frequency-weighted segmental Signal to Noise Ratio (FWSegSNR) \cite{Loizou2011}: Power relation between speech and noise, computed in the frequency domain. Then, larger values indicate better speech quality.
\end{itemize}

\subsection{Speech recognition system}
To measure the speech recognition performance we used the Kaldi recipe for REVERB Challenge \cite{povey2011kaldi, weninger2014merl}. Only the equivalent systems to the Challenge baseline were included, i.e. those based on Gaussian Mixture Model (GMM) and Hidden Markov Models (HMM). 
\begin{itemize}
    \item GMM-HMM: System trained on the clean condition.
    \item GMM-HMM MCT: System trained on the multicondition REVERB dataset.
    \item GMM-HMM MCT + fMLLR: Same as before but including feature space Maximum Likelihood Linear Regression (fMLLR) adaption per utterance.
\end{itemize}

\section{Results and discussion}
\label{sec:res}
\begin{figure}[h!]
\centerline{\includegraphics[width=50mm]{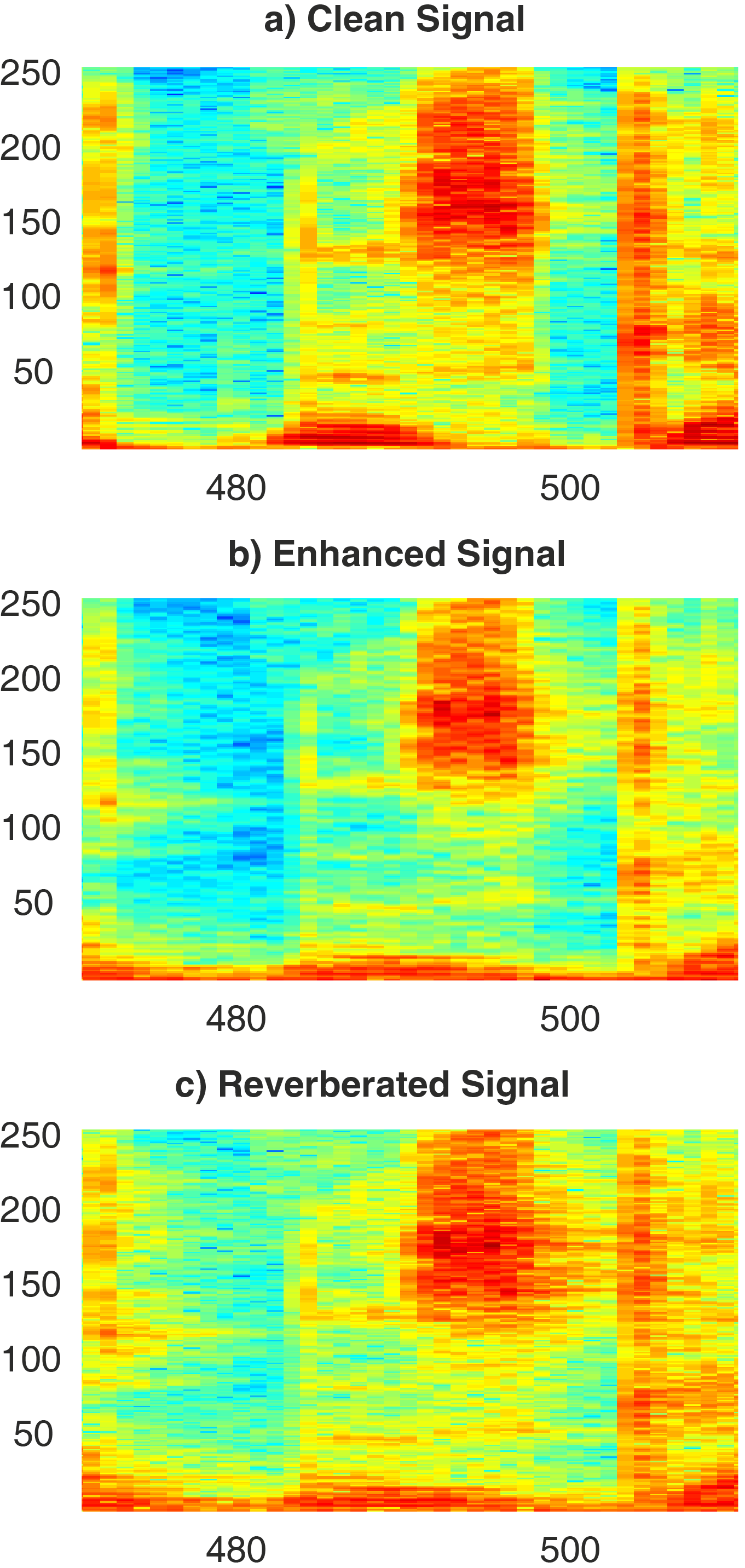}}
\caption{{\it Qualitative example of the method's performance enhancing a reverberated unvoiced sound.}}
\label{fig:goodresult}
\end{figure}

To introduce results, Figure \ref{fig:goodresult} shows a qualitative positive example of the method's performance in a reverberated unvoiced sound. Note, the reverberation effect in the high-frequency area of the spectrum in C) is accurately removed in the enhanced spectrum in B). In the following, results for objective speech quality measures are presented in simulated and real speech data.       

\subsection{Speech quality measures}
Table \ref{resultsrealtest} shows SRMR results for real speech samples from REVERB Development and Evaluation datasets, and for the simulated speech from DIRHA dataset, using the proposed WRN based speech enhancement method. Note that as these datasets do not provide clean references the rest of the quality measures were not computed. Shadowed cells highlight those results where the speech enhancement methods outperformed the baseline. 

The proposal outperforms the baseline on REVERB Evaluation and DIRHA dataset. The performance in different datasets supports the robustness of the method, indicating that its parameters are not adjusted to some specific set of speech signals. Furthermore, note the resulting WRN model, trained with artificial synthesized reverberation, is effective in dealing with reverberated and noisy speech from real-world
scenarios. 

\begin{table} [h]
\caption{\label{resultsrealtest} {\it SRMR for Real REVERB Development, REVERB Evaluation, and DIRHA datasets.}}
\centerline{
\begin{tabular}{|c||c|c|}
\hline
Dataset & baseline & enhanced \\ \hline
REVERB Development & 3.79 & 3.67 \\ \hline  
REVERB Evaluation & 3.18 & \cellcolor{gray!35}3.24 \\ \hline 
DIRHA & 2.29 & \cellcolor{gray!35}2.90 \\ \hline  
\end{tabular}
}
\end{table}

Table \ref{resultstest} presents objective quality measures for simulated speech samples from REVERB Development and Evaluation datasets. There are results using the proposed WRN based speech enhancement method and a statistical speech enhancement method (MM-LSA \cite{Malah1999}), for comparison purposes. Results are consistent in both, REVERB Development and Evaluation datasets, enhanced signals achieved better quality in most of the scenarios evaluated. Moreover, note that results for WRN enhancement method clearly outperform MM-LSA. This fact supports previous theoretical statements about the capacity of neural network solutions for representing the complex structure of noisy and reverberated speech signals, beyond the limitations of the traditional statistical-based approaches.

Anyway, speech from $Room1$ and $near$ speaker-microphone distances was harder to enhance, which indicates the WRN was not trained enough for these scenarios. Due to the lack of exact room size values in the test dataset description, WRN data training (table \ref{tablednntraining}) includes a reasonable estimation of a small room size. However, this is probably not small enough for $Room1$. Furthermore, the distance configuration of the data training design considers the speaker/microphone can be randomly situated all-around the room, modeling it with a uniform data distribution. This left low probability for the specific test data distances of $near$ (50 m) and $far$ (250 m). In order to improve results in these scenarios, the data training in future approaches should include smaller room sizes, as well as change the function to model the speaker-microphone distances, in order to increase the probability in certain distances. However, caution should be taken with too much fitting of the training data to some specific dataset. A better compromise between the network generalization and the test data characteristics will probably be a more reasonable solution. 
\begin{table*} [h!]
\caption{\label{resultstest} {\it Quality measures results in REVERB Development and Evaluation datasets using two speech enhancement methods: the statistical speech estimator MMLSA and the proposed WRN.}}
\centerline{
\begin{tabular}{|c||c|c|c|c|c|c|c||c|c|c|c|c|c|c|}
\hline
 \multirow{3}{*}{Quality measures} & \multicolumn{7}{c||}{REVERB Development dataset} & \multicolumn{7}{c|}{REVERB Evaluation dataset} \\ \cline{2-15}
        & \multicolumn{2}{c|}{Room1} & \multicolumn{2}{c|}{Room2} & \multicolumn{2}{c|}{Room3} & Average & \multicolumn{2}{c|}{Room1} & \multicolumn{2}{c|}{Room2} & \multicolumn{2}{c|}{Room3} & Average   \\ \cline{2-15}
        & Near & Far & Near & Far & Near & Far & - & Near & Far & Near & Far & Near & Far & -  \\
\hline
\multicolumn{15}{|c|}{\textbf{Baseline: Noisy and reverberated speech}} \\
\hline 
CD & 1.96 & 2.65 & 4.58 & 5.08 & 4.20 & 4.82 & 3.88 & 1.99 & 2.67 & 4.63 & 5.21 & 4.38 & 4.96 & 3.97 \\
\hline
LLR & 0.34 & 0.38 & 0.51 & 0.77 & 0.65 & 0.85 & 0.58 & 0.35 & 0.38 & 0.49 & 0.75 & 0.65 & 0.84 & 0.58   \\
\hline
SRMR & 4.37 & 4.63 & 3.67 & 2.94 & 3.66 & 2.76 & 3.67 & 4.50 & 4.58 & 3.74 & 2.97 & 3.57 & 2.73 & 3.68 \\
\hline
FWsegSNR (dB) & 8.10 & 6.75 & 3.07 & 0.53 & 2.32 & 0.14 & 3.48 & 8.12 & 6.68 & 3.35 & 1.04 & 2.27 & 0.24 & 3.62   \\
\hline \hline
\multicolumn{15}{|c|}{\textbf{Enhanced MM-LSA}} \\
\hline 
CD & 4.58 & 4.88 & 5.62 & 6.04 & 5.48 & 5.88 & 5.41 & 4.65 & 4.96 & 5.69 & 6.15 & 5.62 & 6.01 & 5.51   \\
\hline
LLR & 0.81 & 0.88 & 0.98 & 1.14 & 1.07 & 1.20 & 1.01 & 0.83 & 0.90 & 0.97 & 1.14 & 1.08 & 1.21 & 1.02  \\
\hline
SRMR & \cellcolor{gray!35} 4.99 & \cellcolor{gray!35} 5.65 & \cellcolor{gray!35} 4.68 & \cellcolor{gray!35} 4.75 & \cellcolor{gray!35} 4.72 & \cellcolor{gray!35} 4.64 & \cellcolor{gray!35} 4.91 & \cellcolor{gray!35} 5.22 & \cellcolor{gray!35} 5.82 & \cellcolor{gray!35} 4.96 & \cellcolor{gray!35} 4.95 & \cellcolor{gray!35} 4.81 & \cellcolor{gray!35} 4.66 & \cellcolor{gray!35} 5.07 \\
\hline
FWsegSNR (dB) & 5.29 & 4.53 & 2.00 & \cellcolor{gray!35}1.17 & 2.30 & \cellcolor{gray!35} 1.45 & 2.79 & 5.19  & 4.57 & 2.05 & \cellcolor{gray!35} 1.30 & 1.94 & \cellcolor{gray!35} 1.13 & 2.70  \\
\hline \hline
\multicolumn{15}{|c|}{\textbf{Enhanced WRN}} \\
\hline 
CD & 1.99 & \cellcolor{gray!35}2.40 & 4.58 & \cellcolor{gray!35}4.85 & \cellcolor{gray!35}3.99 & \cellcolor{gray!35}4.45 & \cellcolor{gray!35}3.71 & 2.02 & \cellcolor{gray!35}2.43 & \cellcolor{gray!35}4.61 & \cellcolor{gray!35}4.99 & \cellcolor{gray!35}4.15 & \cellcolor{gray!35}4.56 & \cellcolor{gray!35}3.59 \\
\hline
LLR & 0.35 & \cellcolor{gray!35}0.34 & \cellcolor{gray!35}0.48 & \cellcolor{gray!35}0.58 & \cellcolor{gray!35}0.60 & \cellcolor{gray!35}0.68 & \cellcolor{gray!35}0.51 & 0.36 & \cellcolor{gray!35}0.35 & \cellcolor{gray!35}0.46 & \cellcolor{gray!35}0.59 & \cellcolor{gray!35}0.60 & \cellcolor{gray!35}0.67 & \cellcolor{gray!35}0.47 \\
\hline
SRMR & 3.93 & 4.47 & 3.38 & \cellcolor{gray!35}3.29 & 3.42 & \cellcolor{gray!35}2.94 & 3.57 & 4.04 & 4.48 & 3.46 & \cellcolor{gray!35}3.32 & 3.27 & 2.84 & 3.59  \\
\hline
FWsegSNR (dB) & \cellcolor{gray!35}8.26 & \cellcolor{gray!35}7.47 & \cellcolor{gray!35}3.29 & \cellcolor{gray!35}1.38 & \cellcolor{gray!35}2.66 & \cellcolor{gray!35}0.73 & \cellcolor{gray!35}3.96 & \cellcolor{gray!35}8.28 & \cellcolor{gray!35}7.54 & \cellcolor{gray!35}3.57 & \cellcolor{gray!35}1.79 & \cellcolor{gray!35}2.54 & \cellcolor{gray!35}0.88 & \cellcolor{gray!35}4.80 \\
\hline \hline 
\end{tabular}
}
\end{table*}

Note that results for FWsegSNR are outperforming baseline in all the scenarios evaluated, indicating the method was successful at the denoising task, albeit SNR signals level were not especially low. This is a very interesting result, considering that usually, enhancement methods decrease performance at high SNR levels, such as FWsegSNR results for MM-LSA. In this case, as methods cannot find strong noise patterns to remove, they are used to forcing the estimation computing worse enhanced signals than the original noisy one. However, the WRN is not directly denoising the corrupted signal, it is actually constructing a ``clean'' estimation of the corrupted signal from what it learned as clean speech. It takes advantages of the residual connection to hold up a linear shortcut to the input, and use it for adding or subtracting corrections to the non-linear path. Consequently, the WRN performance is less vulnerable to different SNR levels. 

\subsection{Speech recognition}

Table \ref{resultsASR} presents Word Error Rate (WER) results for REVERB Development and Evaluation datasets, using clean, noisy and reverberated speech signals as baselines, as well as the enhanced speech from WRN and a traditional statistical-based enhancement method (MM-LSA). Mostly WRN enhanced speech achieve improved ASR results for scenarios evaluated. Note the use of enhanced speech through MM-LSA cannot outperform the noisy and reverberated baseline. In this case, enhancement actually deteriorates speech signals for recognition purposes, which was a priori suggested by the low performance of LLR and CD quality measures in the previous section. ASR results in scenarios evaluated are consistent with the behavior of the quality metrics in table \ref{resultstest}. $Room1$ and $near$ distances are the more difficult cases in simulated datasets, while REVERB Development is the more difficult dataset for the real speech. Better design of the data training set can contribute to improving results in these scenarios.

\begin{table*} [h!]
\caption{\label{resultsASR} {\it Speech recognition results expressed in WER for REVERB  Development and Evaluation datasets.}}
\centerline{
\begin{tabular}{|c||c|c|c|c|c|c|c||c|c|c|}
\hline
 \multirow{3}{*}{ASR Systems} & \multicolumn{7}{|c||}{SimData} & \multicolumn{3}{|c|}{RealData} \\ \cline{2-11}
        & \multicolumn{2}{|c|}{Room1} & \multicolumn{2}{|c|}{Room2} & \multicolumn{2}{|c|}{Room3} & Average & \multicolumn{2}{|c|}{Room1} & Average   \\ \cline{2-11}
        & Near & Far & Near & Far & Near & Far & - & Near & Far & -   \\
\hline
\multicolumn{11}{|c|}{\textbf{REVERB Development dataset ||  Baseline Clean speech}} \\
\hline 
GMM-HMM & \multicolumn{2}{|c|}{9.29} & \multicolumn{2}{|c|}{9.76} & \multicolumn{2}{|c|}{9.40} & 9.48 & - & - & -   \\
\hline
GMM-HMM MCT &\multicolumn{2}{|c|}{18.24} & \multicolumn{2}{|c|}{19.67} & \multicolumn{2}{|c|}{19.02} & 18.98 & - & - & -   \\
\hline
GMM-HMM MCT + fMLLR & \multicolumn{2}{|c|}{12.78} & \multicolumn{2}{|c|}{12.1} &\multicolumn{2}{|c|}{12.76} & 12.55 & - & - & -   \\
\hline \hline
\multicolumn{11}{|c|}{\textbf{Baseline: Noisy and reverberated speech}} \\
\hline 
GMM-HMM & 12.78 & 22.25 & 43.36 & 88.19 & 49.53 & 90.68 & 51.13 & 89.58 & 89.61 &  89.59  \\
\hline
GMM-HMM MCT & 14.43 & 17.01 & 21.99 & 45.03 & 25.42 & 50.12 & 29.00 & 56.96 & 52.97 & 54.96   \\
\hline
GMM-HMM MCT + fMLLR & 12.95 & 16.27 & 17.72 & 36.31 & 20.05 & 39.42 & 23.79 & 48.72 & 45.86 & 47.29   \\
\hline \hline
\multicolumn{11}{|c|}{\textbf{Enhanced MM-LSA}} \\
\hline 
GMM-HMM & 31.00 & 49.83 & 64.83 & 91.15 & 67.09 & 91.12 & 65.84 & - & - & -   \\
\hline
GMM-HMM MCT & 24.53 & 31.27 & 36.75 & 63.20 & 39.49 & 61.50 & 42.79 & - & - & -  \\
\hline
GMM-HMM MCT + fMLLR &  20.38 & 34.36 & 30.17 & 52.21 & 31.65 & 52.94 & 35.28 & - & - & -   \\
\hline \hline
\multicolumn{11}{|c|}{\textbf{Enhanced WRN}} \\
\hline 
GMM-HMM & 13.64 & \cellcolor{gray!35}19.89 & \cellcolor{gray!35}41.43 & \cellcolor{gray!35}72.66 & \cellcolor{gray!35}46.76 & \cellcolor{gray!35}80.96 & \cellcolor{gray!35}45.89 & \cellcolor{gray!35}87.27 & \cellcolor{gray!35}86.94 & \cellcolor{gray!35}87.10   \\
\hline
GMM-HMM MCT & \cellcolor{gray!35}14.16 & \cellcolor{gray!35}15.95 & 22.23 & \cellcolor{gray!35}35.94 & \cellcolor{gray!35}24.78 & \cellcolor{gray!35}40.55 & \cellcolor{gray!35}25.60 & \cellcolor{gray!35}56.89 & 55.50 & 56.19  \\
\hline
GMM-HMM MCT + fMLLR &  \cellcolor{gray!35}12.78 & \cellcolor{gray!35}15.29 & 19.40 & \cellcolor{gray!35}28.79 & 20.90 & \cellcolor{gray!35}32.34 & \cellcolor{gray!35}21.58 & 49.34 & 46.55 & 47.94   \\
\hline \hline
\multicolumn{11}{|c|}{\textbf{REVERB Evaluation dataset  ||   Baseline: Clean speech}} \\
\hline 
GMM-HMM & \multicolumn{2}{|c|}{10.60} & \multicolumn{2}{|c|}{10.30} & \multicolumn{2}{|c|}{10.54} & 10.48 & - & - & -   \\
\hline
GMM-HMM MCT & \multicolumn{2}{|c|}{23.04} & \multicolumn{2}{|c|}{22.34} & \multicolumn{2}{|c|}{21.84} & 22.41 & - & - & -   \\
\hline
GMM-HMM MCT + fMLLR &\multicolumn{2}{|c|}{14.18} & \multicolumn{2}{|c|}{13.62} & \multicolumn{2}{|c|}{13.83} & 13.88 & - & - & -   \\
\hline \hline
\multicolumn{11}{|c|}{\textbf{Baseline: Noisy and reverberated speech}} \\
\hline 
GMM-HMM & 14.79 & 23.50 & 40.35 & 81.92 & 51.45 & 88.74 & 50.12 & 89.05 & 87.78 & 88.41   \\
\hline
GMM-HMM MCT & 17.38 & 19.01 & 21.23 & 38.42 & 25.91 & 47.00 & 28.16 & 56.02 & 52.80 & 54.41   \\
\hline
GMM-HMM MCT + fMLLR & 14.09 & 15.91 & 18.68 & 31.07 & 21.37 & 38.60 & 23.29 & 47.72 & 45.86 & 47.29   \\
\hline \hline
\multicolumn{11}{|c|}{\textbf{Enhanced MM-LSA}} \\
\hline 
GMM-HMM & 32.44 & 50.70 & 61.34 & 89.07 & 68.46 & 90.40 & 65.40 & - & - & -   \\
\hline
GMM-HMM MCT & 29.76 & 35.44 & 35.86 & 58.38 & 41.30 & 62.67 & 43.90 & - & - & -   \\
\hline
GMM-HMM MCT + fMLLR & 20.24 & 24.48 & 27.62 & 45.76 & 33.37 & 52.63 & 34.02 & - & - & -   \\
\hline \hline
\multicolumn{11}{|c|}{\textbf{Enhanced WRN}} \\
\hline 
GMM-HMM & 15.35 & \cellcolor{gray!35}18.84 & \cellcolor{gray!35}38.53 & \cellcolor{gray!35}66.31 & \cellcolor{gray!35}47.54 & \cellcolor{gray!35}78.25 & \cellcolor{gray!35}44.14 & \cellcolor{gray!35}85.95 & \cellcolor{gray!35}82.31 & \cellcolor{gray!35}84.13   \\
\hline
GMM-HMM MCT & 18.47 & 19.21 & 21.51 & \cellcolor{gray!35}32.29 & \cellcolor{gray!35}24.94 & \cellcolor{gray!35}38.98 & \cellcolor{gray!35}25.90 & \cellcolor{gray!35}55.86 & \cellcolor{gray!35}52.40 & \cellcolor{gray!35}54.13  \\
\hline
GMM-HMM MCT + fMLLR &  14.92 & \cellcolor{gray!35}15.79 & \cellcolor{gray!35}18.63 & \cellcolor{gray!35}26.70 & 21.85 & \cellcolor{gray!35}33.54 & \cellcolor{gray!35}21.90 & \cellcolor{gray!35}46.98 & 46.73 & \cellcolor{gray!35}46.85   \\
\hline
\end{tabular}
}
\end{table*}

\section{Conclusions and future work}
\label{sec:conc}
This paper has introduced a novel speech enhancement method based on a WRN architecture that takes advantage of the powerful representations obtained from a wide topology of CNN with residual connections. Enhanced speech obtained from the WRN was successfully tested with speech quality metrics and ASR system performance, showing the potentiality of WRN in providing enhanced speech either for intelligibility as for recognition purposes. This is an outstanding result considering that traditional statistical-based methods, e.g. MM-LSA used for comparison purposes, are usually limited in this aim. Best results were obtained for far-field reverberated and noisy speech in three different room sizes. Furthermore, the proposal succeeded in the denoising task for high SNR levels. This is a usual handicap of the enhancement methods, which can effectively enhance very noisy speech but are limited for speech reconstruction with low noise levels. The residual mechanism was extremely useful in this case, since the signal has always a linear shortcut and the non-linear path enhances it in certain steps by adding or subtracting corrections. In practical applications, this is a valuable property because realistic scenarios could challenge the system with many different conditions, and sometimes the real world is not so noisy as research studies consider in experimental setups \cite{Ribas2016}.

Despite results are encouraging, yet the proposal can be further improved. Future work will focus on fine tuning the data training configuration with a view to updating the compromise between generalization and accuracy. We also plan to expand the experimental setup with speech data in alternative scenarios from other datasets. On the other side, the inclusion of perceptual features in the network cost function will be explored in order to improve the performance in the speech reconstruction process. 

\section*{Acknowledgment}
This work has been supported by the Spanish Ministry of Economy and Competitiveness and the European Social Fund through the project TIN2017-85854-C4-1-R. We gratefully acknowledge the support of NVIDIA Corporation with the donation of the Titan Xp GPU used for this research. This material is based upon work supported by Google Cloud.

\bibliographystyle{IEEEbib}


\end{document}